\documentclass[runningheads]{llncs}
\usepackage[T1]{fontenc}

\usepackage[T1]{fontenc}
\usepackage[utf8]{inputenc} 
\DeclareUnicodeCharacter{03C4}{\ensuremath{\tau}} 

\DeclareUnicodeCharacter{2014}{\textemdash}
\DeclareUnicodeCharacter{2013}{\textendash}
\DeclareUnicodeCharacter{2019}{'}  
\usepackage{url} 

\usepackage{booktabs}   
\usepackage{multirow}   
\usepackage{booktabs}

\usepackage{subcaption} 
\usepackage{amsmath}
\usepackage{graphicx,verbatim}
\usepackage{bbding}      
\usepackage{placeins}   

\usepackage{array} 
\newcolumntype{+}{!{\vrule width 1pt}}

\begin{document}
\title{Fitzpatrick Thresholding for Skin Image Segmentation}

\author{
    Duncan Stothers \inst{1} \orcidID{0000-0001-6873-851X}\Envelope
\and
    Sophia Xu \inst{3} \and
    Carlie Reeves \inst{4}
 \and
    Lia Gracey \inst{2} \orcidID{0009-0007-6037-2518} 
}

\authorrunning{D. Stothers et al.}

\institute{
   Independent Researcher, San Francisco, CA, USA
  \email{duncanstothers@alumni.harvard.edu}
  \and
  Vagelos College of Physicians and Surgeons,
           Columbia University, New York, NY, USA
  \email{sx2400@cumc.columbia.edu}
  \and
  University of Mississippi Medical Center, Jackson, MS, USA
  \email{CReeves4@umc.edu}
  \and
  Division of Dermatology, The University of Texas at Austin, Dell Medical School, Austin, Texas
  \email{lia.gracey@austin.utexas.edu}
}

\maketitle              
\begin{abstract}
Accurate estimation of the body surface area (BSA) involved by a rash, such as psoriasis, is critical for assessing rash severity, selecting an initial treatment regimen, and following clinical treatment response. Attempts at segmentation of inflammatory skin disease such as psoriasis  perform markedly worse on darker skin tones, potentially impeding equitable care. We assembled a psoriasis dataset sourced from six public atlases, annotated for Fitzpatrick skin type, and added detailed segmentation masks for every image. Reference models based on  U‑Net, ResU-Net, and SETR-small are trained without tone information. On the tuning split we sweep decision thresholds and select (i) global optima and (ii) per Fitzpatrick skin tone optima for Dice and binary IoU.  Adapting Fitzpatrick specific thresholds lifted segmentation performance for the darkest subgroup (Fitz VI) by up to +31 \% bIoU and +24 \% Dice on UNet, with consistent, though smaller, gains in the same direction for ResU-Net (+25 \% bIoU, +18 \% Dice) and SETR-small (+17 \% bIoU, +11 \% Dice). Because Fitzpatrick skin tone classifiers trained on Fitzpatrick‑17k now exceed 95 \%\ accuracy, the cost of skin tone labeling required for this technique has fallen dramatically. Fitzpatrick thresholding is simple, model‑agnostic, requires no architectural changes, no re-training, and is virtually cost free. We demonstrate the inclusion of Fitzpatrick thresholding as a potential future fairness baseline.
\end{abstract}


%



\section{Background}

\subsection{Significance}
Skin rashes remain one of the most frequent reasons for new primary‑care
encounters, accounting for more than 13 million office visits annually in the United States and rising \cite{CDC_NAMCS_NHAMCS_WebTables_2023}.  Diagnostic accuracy is unevenly distributed: both practicing dermatologists and trainees perform noticeably worse on images of darker skin tones \cite{Daneshjou2022_DermAI_Disparities,Diao2022_DarkSkinChallenge}. The accurate assessment of body surface area (BSA) affected by skin conditions, such as rashes, is crucial for clinical decision-making. Yet, physicians still rely on the outdated “1 palm = 1 percent BSA” method where BSA involved with a rash is estimated using the patient’s palm size. This subjective measurement can lead to under- or over-treatment in the clinic. Additionally, a minimum threshold of BSA involvement is a criterion for payors in insurance coverage decisions, which makes accurate calculations imperative for a patient to be eligible for more advanced biologic treatments and for following treatment response. More specifically, BSA is an important calculation in the widely used Psoriasis Area and Severity Index that is most often deployed in clinical trial settings to assess baseline and treatment response for new therapeutics; these common measures are subjective and prone to human error \cite{Bozek2017Reliability}. No widely used tools exist to automate these important assessments in all skin types\cite{Mogawer2020_VES_VASI,Silverberg2021_IGA_BSA,Yoo2022_ImageJ_BSA}.
Any systematic error in segmenting lesions on dark skin therefore propagates directly into PASI scores, treatment eligibility, and ultimately patient outcomes.

\subsection{Previous Work}

\subsubsection{Early ISIC Analyses Highlight Tone Bias}
The first wave of ISIC challenge papers demonstrated that convolutional
networks trained almost exclusively on Fitzpatrick I–III images attained
dermatologist‑level accuracy on similarly light‑skinned test sets, yet their performance degraded noticeably on darker tones \cite{Kinyanjui2020Fairness,Groh2021Fitzpatrick17k}. Follow‑up studies on ISIC 2018, Fitzpatrick‑17k, and DDI quantified AUROC and sensitivity gaps of 10–35 pp favoring light skin \cite{Kinyanjui2020Fairness,Groh2021Fitzpatrick17k,Daneshjou2022Disparities}. The consensus emerging from
this literature is that distributional shift in pigmentation, not just lesion morphology, drives a substantial share of the error.

\subsubsection{From Complex Debiasing Schemes to Stratified Operating Points}
Most responses to the documented bias have focused on sophisticated data‑ or model‑centric fixes—balanced resampling, adversarial representation learning, group‑adaptive batch normalization, or fairness‑guided pruning
\cite{Pakzad2022CIRCLe,Xu2023FairAdaBN,Wu2023FairPrune}.  A conceptually simpler alternative, rooted in the equalized‑odds post‑processing of Hardt's 2016 approach \cite{Hardt2016Equality}, is to select a \emph{separate decision threshold} for each Fitzpatrick group so that error rates align across tones.

\subsubsection{The FPR–TPR Trade‑off in Binary Classification}
Applying stratified thresholds to binary classification is not trivial:
raising sensitivity for an under‑served group often worsens its false‑positive rate, and—by impossibility results—one cannot simultaneously satisfy perfect calibration and equalized odds once prevalence differs \cite{Kleinberg2017TradeOffs,Pleiss2017FairnessCalibration}.  Consequently, dermatology researchers have tended to pursue fairness during training \cite{Pakzad2022CIRCLe,Xu2023FairAdaBN,Wu2023FairPrune}, where the utility–equity trade‑off is perceived as more controllable, rather than post‑hoc calibration.

\subsubsection{Segmentation: A Setting Where Tone‑Specific Optima Exist}
Segmentation changes the landscape.  Each image yields a dense probability
map, and there is, in principle, a threshold that maximizes Dice or bIoU for every subgroup.  If the score distributions for Fitzpatrick V–VI are shifted left—as empirical histograms repeatedly show \cite{Bencevic2024SkinColorBias} —a universal cut‑off under‑segments dark skin.  Calibrating per‑tone thresholds can therefore improve both subgroup Dice and \emph{overall} performance, because each group operates closer to its own theoretical optimum. This observation motivates the present study, which evaluates Fitzpatrick‑specific thresholding in the clinically consequential task of psoriasis BSA estimation.

\subsection{Clinical Relevance of Precise BSA Estimation}
Psoriasis management provides an ideal test‑bed for tone‑aware segmentation because small changes in the BSA assessment directly translate to different treatment pathways. The PASI scoring rubric weights percent‑involved BSA in each anatomical region; a 5–10 percentage‑point error may erroneously move a patient into a different disease severity category. In a recent review of machine learning BSA estimators, skin tone discussion was omitted from all segmentation approaches \cite{Li2024_psoriasis_review}, with the sole exception that in one study it was shown that error modes exist where healthy darker skin regions are sometimes mis‑classified as lesional \cite{George2017_psoriasis_segmentation}. Demonstrating that Fitzpatrick‑specific thresholding can reduce this bias would offer a pragmatic, model‑agnostic fairness intervention with potential uses both in clinical trials and photo-based tele-dermatology.

\section{Methodology}

\subsection{Data Collection}

We assembled a large publicly available psoriasis dataset by
sourcing from six open dermatology repositories: Derm Atlas Brazil \cite{DermAtlasBrazil2024},
DermIS \cite{DermIS}, DermNet NZ \cite{DermNetNZ}, the Hellenic
Dermatology Atlas \cite{HellenicDermAtlas}, the Interactive Dermatology Atlas \cite{InteractiveDermAtlas}, and Fitzpatrick‑17k \cite{Groh2021Fitzpatrick17k}. Subtypes of psoriasis that were excluded included pustular variants and isolated nail disease, filtered out by keyword rules and manual dermatologist review. Duplicates were removed, and patient IDs were assigned to prevent leakage between train, tune, and test sets. The final dataset contained 754 psoriasis images from 631 patients.

\subsection{Skin‑Tone Annotation and Segmentation Labels}

Each retained image was independently labeled with a Fitzpatrick type
(I–VI) by a board‑certified dermatologist. Pixel‑level diseased‑skin masks were produced using the VIA Image Annotator tool \cite{Dutta2019VIA}. Three assistants (medical student, resident physician, and graduate research assistant) drew initial polygon masks; a board‑certified MD–PhD dermatologist specializing in psoriasis revised every mask to ensure high quality segmentation masks, especially on difficult cases such as low contrast lesions on darker skin tones.

\begin{figure}
\includegraphics[width=\textwidth]{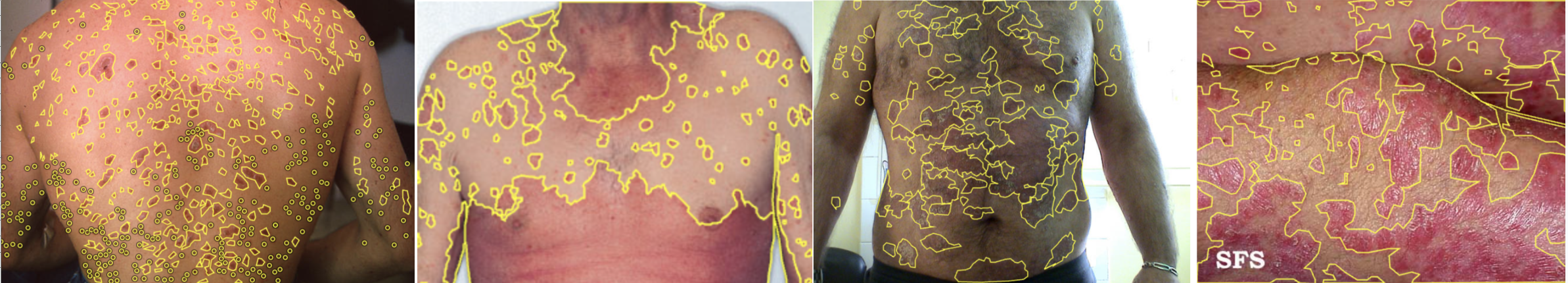}
\caption{Examples of high detail manual skin-disease labeling employed in the study.} 
\end{figure}

\begin{table}[!t]
\caption{Per-dataset image counts by Fitzpatrick skin-type}
\label{tab:fitz_counts}
\centering
\begin{tabular}{|l|r|r|r|r|r|r|r|}
\hline
\textbf{Dataset} & \textbf{I} & \textbf{II} & \textbf{III} & \textbf{IV} & \textbf{V} & \textbf{VI} & \textbf{Total} \\ \hline
Brazil            &  0 &   2 &  29 &  40 &  40 &  4 & 115 \\
DermIS            & 22 &  67 &   2 &   3 &   3 &  0 &  97 \\
DermNetNZ         & 27 & 171 &  68 &  32 &   7 &  5 & 310 \\
Fitzpatrick\,17k  & 35 &  57 &   8 &  11 &   7 &  1 & 119 \\
Hellenic          &  1 &  29 &  16 &   6 &   0 &  0 &  52 \\
Interactive       &  0 &  24 &  22 &   4 &   5 &  6 &  61 \\ \hline
\textbf{Column total} & \textbf{85} & \textbf{350} & \textbf{145} & \textbf{96} & \textbf{62} & \textbf{16} & \textbf{754} \\ \hline
\end{tabular}
\end{table}

\subsection{Data Split}

Patient‐level IDs were stratified by Fitzpatrick skin tone and, within each stratum, randomly permuted with a fixed seed (0).  Stratified samples were then allocated to the training, tuning, and held-out test sets in a 30 / 30 / 40 proportion, ensuring balanced skin-tone representation and complete patient independence across partitions.

\subsection{Model Architecture and Training Protocol}

We benchmark three architectures chosen to represent successive stages in semantic segmentation design while remaining practical for a single‑GPU medical study. U‑Net \cite{ronneberger2015unet} is the canonical encoder–decoder CNN against which most dermatology work is still compared. Our 256 × 256 implementation (four down‑sampling stages, two 3 × 3 convs per block, batch‑norm everywhere) contains 31.1 trainable parameters and therefore serves as a strong, yet widely recognizable, baseline. Residual U‑Net \cite{Zhang2018ResUNet} keeps the same overall topology and feature widths but replaces each plain block with a pre‑activation residual pair plus a squeeze‑and‑excite (SE) channel attention gate. These lightweight additions raise the capacity only marginally to 33.1 M parameters, letting us test whether better optimization and local attention alone can reduce skin‑tone bias. Finally, a 21M parameter reference implementation of SETR-small \cite{Zheng2021SETR} swaps the convolutional encoder for a ViT‑S/16 backbone (12 transformer layers, 6 heads, 384‑D embeddings; positional tokens only) followed by a one‑layer up sampling head. Because the encoder is fully self‑attentional and translation‑equivariant only after training, SETR probes whether long‑range context helps fairness on our limited psoriasis corpus. All three networks are trained with identical 256 × 256 inputs, vanilla SGD optimization with 0.9 momentum, learning rates of 0.01 for UNet and ResUNet (0.0004 for SETR for stability), an identical simple flip, rotate, and jitter data augmentation scheme, early‑stopping watching the validation bIoU with a patience setting of 15, and an identical 3:1 weighted binary cross entropy + dice loss; thus any performance differences are most likely attributed to (i) architectural choice and (ii) the use of Fitz‑specific versus global operating points, rather than to confounding hyper‑parameter changes.

\subsection{Operating‑Point Search}

To evaluate threshold sensitivity we swept decision cut‑offs
\(\tau\in[0.001,0.99]\) in steps of 0.001 on the \emph{tuning} split and
computed binary Intersection‑over‑Union (bIoU) and Dice:

\[
\mathrm{Dice}(\tau)=
\frac{2\,|\hat{M}(\tau)\cap M|}{|\hat{M}(\tau)|+|M|}, \quad
\mathrm{bIoU}(\tau)=
\frac{|\hat{M}(\tau)\cap M|}{|\hat{M}(\tau)\cup M|},
\]

where \(M\) is the ground‑truth mask and
\(\hat{M}(\tau)=\{\,p\ge\tau\,\}\).
We recorded:
\begin{itemize}
  \item Two \emph{overall} optima, \(\tau^{\mathrm{Dice}}_\mathrm{all}\) and
        \(\tau^{\mathrm{bIoU}}_\mathrm{all}\), maximizing performance across
        the entire tuning set.
  \item Twelve \emph{tone‑stratified} optima,
        \(\tau^{\mathrm{Dice}}_g,\tau^{\mathrm{bIoU}}_g\) for
        \(g\in\{\text{I},\dots,\text{VI}\}\), each maximizing the metric
        within its Fitzpatrick subgroup.
\end{itemize}
All operating points were then frozen and evaluated once on the unseen test
set to quantify gains from tone‑specific calibration.

\section{Results}

\begin{figure}
\includegraphics[width=\textwidth]{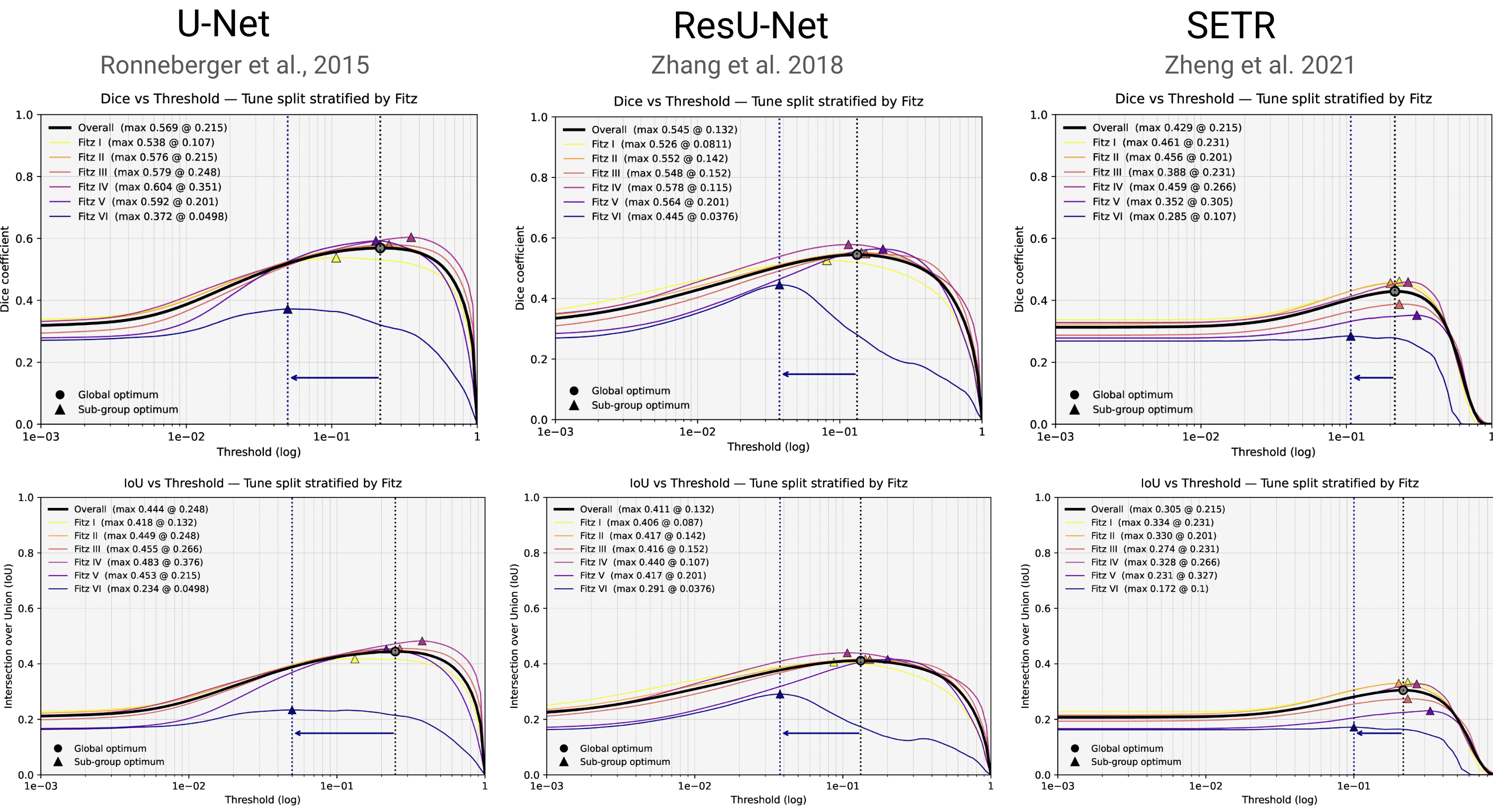}
\caption{Data plotted from the validation set. Arrows highlight that the optimal operating points for Fitz VI is consistently and significantly lower than for other Fitzpatrick tones. This observation is consistent between architectures: from all-conv UNet to all-attention SETR, as well as between metrics Dice (top row) and bIoU (bottom row). Fitz tones I-V have optimal operating points around the aggregate overall optimum which is shown in black.} \label{fig1}
\end{figure}

\begin{figure}[]
\centering  
\includegraphics[width=1.0\textwidth]{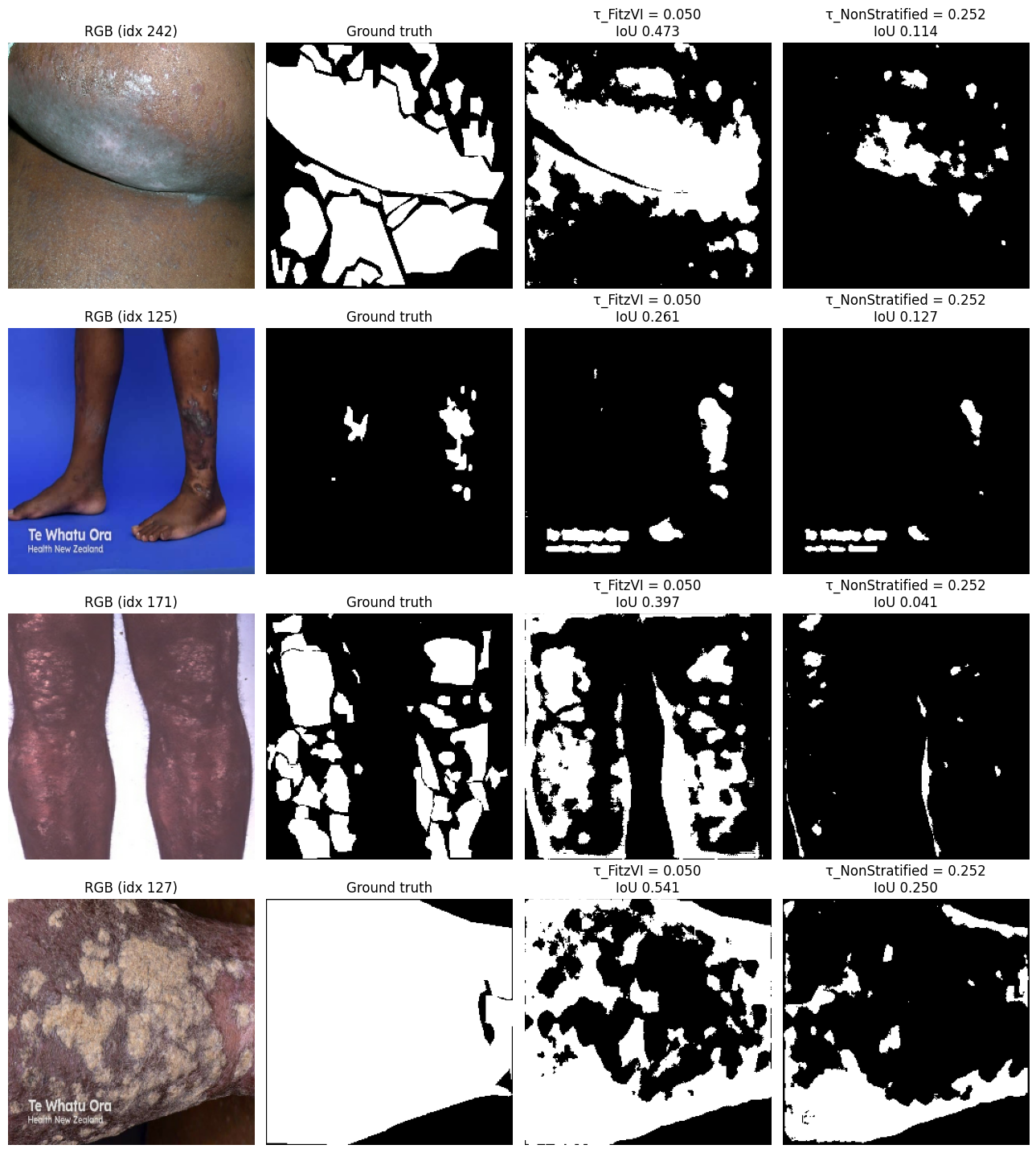}
\caption{The lower Fitz-VI optimized threshold (column 3) captures much more of the diseased skin than the global optimized operating point (column 4). Inference examples from U-Net.} 
\end{figure}

\begin{table}[!ht]
  \caption{Segmentation performance by skin-tone subset.  
           $\tau_g$: global threshold; $\tau_{\mathrm{F}}$: Fitz-specific threshold.}
  \label{tab:fitz_three_models_ordered}
  \centering
  \resizebox{\textwidth}{!}{%
    \begin{tabular}{ll*{3}{ccc}}
      \toprule
      & &
      \multicolumn{3}{c}{\textbf{U-Net}} &
      \multicolumn{3}{c}{\textbf{ResU-Net}} &
      \multicolumn{3}{c}{\textbf{SETR-small}}\\
      \cmidrule(lr){3-5}\cmidrule(lr){6-8}\cmidrule(lr){9-11}
      \textbf{Metric} & \textbf{Subset} &
      $\tau_g$ & $\tau_\mathrm{F}$ & $\Delta$ (\%)  &
      $\tau_g$ & $\tau_\mathrm{F}$ & $\Delta$ (\%)  &
      $\tau_g$ & $\tau_\mathrm{F}$ & $\Delta$ (\%) \\
      \midrule
      \multirow{7}{*}{Dice}
       & Overall & 0.682 & —    & —      & 0.647 & —    & —      & 0.510 & —    & —\\
       & Fitz I  & 0.569 & 0.575 & $+0.97$ & 0.543 & 0.556 & $+2.41$ & 0.472 & 0.470 & $-0.54$\\
       & Fitz II & 0.584 & 0.584 & $0.00$  & 0.587 & 0.585 & $-0.36$ & 0.512 & 0.516 & $+0.66$\\
       & Fitz III& 0.649 & 0.650 & $+0.05$ & 0.619 & 0.622 & $+0.49$ & 0.474 & 0.475 & $+0.26$\\
       & Fitz IV & 0.691 & 0.657 & $-4.94$ & 0.723 & 0.730 & $+1.03$ & 0.597 & 0.575 & $-3.71$\\
       & Fitz V  & 0.557 & 0.563 & $+1.16$ & 0.593 & 0.561 & $-5.26$ & 0.449 & 0.442 & $-1.52$\\
       & Fitz VI & 0.475 & 0.590 & $(+24.13)$& 0.556 & 0.656 & $(+18.01)$& 0.535 & 0.594 & $(+11.04)$\\
      \midrule
      \multirow{7}{*}{bIoU}
       & Overall & 0.558 & —    & —      & 0.514 & —    & —      & 0.371 & —    & —\\
       & Fitz I  & 0.424 & 0.434 & $+2.42$ & 0.398 & 0.411 & $+3.24$ & 0.338 & 0.336 & $-0.55$\\
       & Fitz II & 0.457 & 0.457 & $0.00$  & 0.454 & 0.452 & $-0.42$ & 0.373 & 0.376 & $+0.92$\\
       & Fitz III& 0.514 & 0.514 & $+0.01$ & 0.478 & 0.480 & $+0.45$ & 0.335 & 0.338 & $+0.72$\\
       & Fitz IV & 0.564 & 0.530 & $-6.06$ & 0.600 & 0.613 & $+2.31$ & 0.456 & 0.438 & $-4.05$\\
       & Fitz V  & 0.414 & 0.424 & $+2.37$ & 0.455 & 0.428 & $-6.00$ & 0.311 & 0.306 & $-1.74$\\
       & Fitz VI & 0.353 & 0.464 & $(+31.46)$& 0.423 & 0.527 & $(+24.63)$& 0.395 & 0.463 & $(+17.14)$\\
      \bottomrule
    \end{tabular}%
  }
\end{table}

\subsubsection{Visual confirmation.}
Figure~\ref{fig1} plots validation bIoU and dice versus threshold for each tone.  The curves illustrate a consistent left‑shift for Fitzpatrick VI, explaining why the universal cut‑off under‑segments darker skin. The arrows mark the tone‑specific optima chosen during calibration; note that lighter tones cluster around the global optimum, whereas tone VI requires substantially lower thresholds to maximize Dice. 

Figure 3 illustrates example Fitz VI images that visually illustrate how the lower Fitz VI optimized operating points captures significantly more of the diseased skin than the globally optimized operating point which is dominated by lighter skin tones. 

\subsubsection{Quantitative improvements of Fitzpatrick thresholding}
In Table~\ref{tab:fitz_three_models_ordered} we can see applying a single global threshold already yields reasonable performance for all three networks, but re-tuning the operating point for each Fitzpatrick subgroup uncovers systematic gains that disproportionately benefit the darkest skin tones. Because the Fitzpatrick specific operating points are only exercised within its own subgroup, the macro-average for both dice and bIoU across all tones moves minimally for every architecture.

\section{Discussion}

\subsubsection{Labor intensive skin‑tone annotation to verify equitable stratified performance is no longer a bottleneck.}
Historically, per‑image Fitzpatrick labels required labour‑intensive, subjective
grading by board‑certified dermatologists—prohibitive for million‑image
repositories. The advent of high‑accuracy tone classifiers trained on Fitzpatrick‑17k that reach
\(>\!95\%\) balanced accuracy in external validation
 across a broad cross section of dermatology diseases \cite{Groh2021Fitzpatrick17k}. In practice, a lightweight classifier adds minimal inference time and can be applied retrospectively to every archive or even prospectively on‑device.

\subsubsection{A practical addition to the fairness toolbox.}
Per‑group threshold calibration is an immediately deployable fairness lever—orthogonal to, and composable with, data balancing, representation alignment, FairAdaBN \cite{Xu2023FairAdaBN}, or FairPrune \cite{Wu2023FairPrune}.  We therefore recommend that future skin‑segmentation studies:
\begin{enumerate}
  \item Define in the metadata the skin tone of images in the pre-processing pipeline via an automated method such as a Fitzpatrick17k classifier.
  \item Tune τg on a validation split, or ideally on the set of predictions from a cross-fold validation.
  \item Report both overall and tone‑stratified metrics at those
        thresholds.
\end{enumerate}
Either the stratified performance with the global threshold is equitable across skin tones, or it may not be, in which case Fitzpatrick thresholding provides a lever to lower the under performance as demonstrated here in psoriasis segmentation. Doing so requires no architectural change, no re‑training, and negligible runtime overhead, yet—as shown here—can significantly increase performance on the darkest skin tones.  Given its simplicity and efficacy, Fitzpatrick‑specific thresholding could become a standard baseline for ISIC fairness tracks and for any clinical deployment of dermatology segmentation models.

    

\begin{credits}
\subsubsection{\ackname} No sponsoring company was involved in the production of this work.

\subsubsection{\discintname}
The authors have no competing interests to declare that are relevant to the content of this article.
\end{credits}

%
%
%
\newpage

\bibliographystyle{splncs04}
\bibliography{bibliography}
\end{document}